\documentclass{iau}

\usepackage{graphicx}
\usepackage{amsmath}
\usepackage{natbib}
\usepackage{hyperref}
\usepackage{orcidlink}
\usepackage{multirow}
\usepackage{subcaption}

\begin{document}

\lefttitle{Ira Sharma\orcidlink{0009-0009-0602-0751} \textit{et al.}}
\righttitle{Tidal Tails and Their Dynamics in Open
Clusters Using Gaia DR3}
\jnlPage{1}{7} 
% Journal info
\jnlDoiYr{2025}
\doival{10.1017/xxxxx}

\aopheadtitle{Proceedings of the IAU Symposium}
% Editor names as per IAU style
\editors{C. Sterken, J. Hearnshaw \& D. Valls-Gabaud, eds.}

% Title
\title{Tidal Tails and Their Dynamics in Open Clusters Using Gaia DR3}

% Authors
\author{
    Ira Sharma$^{1,2}$\orcidlink{0009-0009-0602-0751},
    Vikrant Jadhav$^3$\orcidlink{0000-0002-8672-3300},
    Annapurni Subramaniam$^1$\orcidlink{0000-0003-4612-620X}
}

\affiliation{
    $^1$Indian Institute of Astrophysics, 2nd Block, Koramangala, Bangalore-560034, India\\
    email: \email{irasharma2208@gmail.com}\\
    $^2$IISER Mohali, Knowledge City, Sector 81, SAS Nagar, Manauli PO 140306, Punjab, India\\
    $^3$Helmholtz-Institut für Strahlen- und Kernphysik, Universität Bonn, Nussallee 14-16, 53115 Bonn, Germany
}

% \affiliation{Department of Physical Sciences, IISER Mohali, India\\
% \email{irasharma2208@gmail.com}}

\begin{abstract}
This research presents unsupervised machine learning and statistical techniques to detect and analyze tidal tails in five open clusters,
BH 164, Alessi 2, NGC 2281, NGC 2354, and M67, with ages ranging from 60 Myr to 4 Gyr, using Gaia DR3 data. Color-magnitude diagram (CMD) matching, principal component analysis (PCA), and density-based clustering were used to detect the tails.  Tidal tails were found in all clusters, extending 40 to 100 pc and containing 100 to 200 stars. Other statistical methods were used to study the stellar kinematics, photometry, and spatial distribution of the detected features. The tails had luminosity functions similar to their parent clusters, generally lacked massive stars, and showed higher binary fractions. Significant rotation was detected in M67 and NGC 2281 for the first time.
\end{abstract}


\maketitle

\section{Introduction}

Tidal tails are extended stellar streams stretching from clusters due to the MW's gravitational effects, providing information about a cluster's dynamical state, trajectory, and mass loss history. Historically, the detection of open cluster tidal tails has been challenging due to low stellar densities and significant field star contamination from the Galactic disc. However, the Gaia mission, particularly Gaia DR2 \citep{2018A&A...616A...1G} and Gaia EDR3 \citep{2021A&A...649A...1G}, with unprecedented astrometric accuracy, has greatly improved the detection of such extended features. 

This study used Gaia DR3 \citep{2023A&A...674A...1G} to identify and examine tidal tails in five open clusters: BH 164, NGC 2354, Alessi 2, NGC 2281, and M67, chosen based on prior evidence of extended tidal structures \citep{2022MNRAS.517.3525B, Tarricq2022A&A...659A..59T, Hunt2023A&A...673A.114H, 2024A&A...686A..42H, 2020PASJ...72...47G}. The clusters span distances of 400--1300 pc and ages of 60 Myr--4 Gyr, which makes for a wide range of Galactic environments. We developed an optimized methodology to detect tidal tails in distant systems by combining clustering algorithms, kinematic analysis, and validation techniques, achieving significant improvements over earlier studies. The detailed research work is presented in \citet{2025arXiv250819457S}.
\section{Cluster Identification and Tidal Tail Detection}

\subsection{Data Filtering and Preparation} 
Proper motion (PM) cuts were applied using cluster-specific bounds based on the literature values of PM distributions. Stars with absolute PM errors $< 0.5$ mas/yr, positive parallaxes, and relative parallax errors $\sigma_{\varpi}/\varpi < 0.1$ formed the \emph{Filtered Sample}. The identification of the core cluster used five normalized astrometric features: \texttt{ra}, \texttt{dec}, \texttt{pmra}, \texttt{pmdec}, and \texttt{distance}.

\begin{figure}[t]
\centering
\includegraphics[width=0.82\textwidth, height=0.39\textheight]{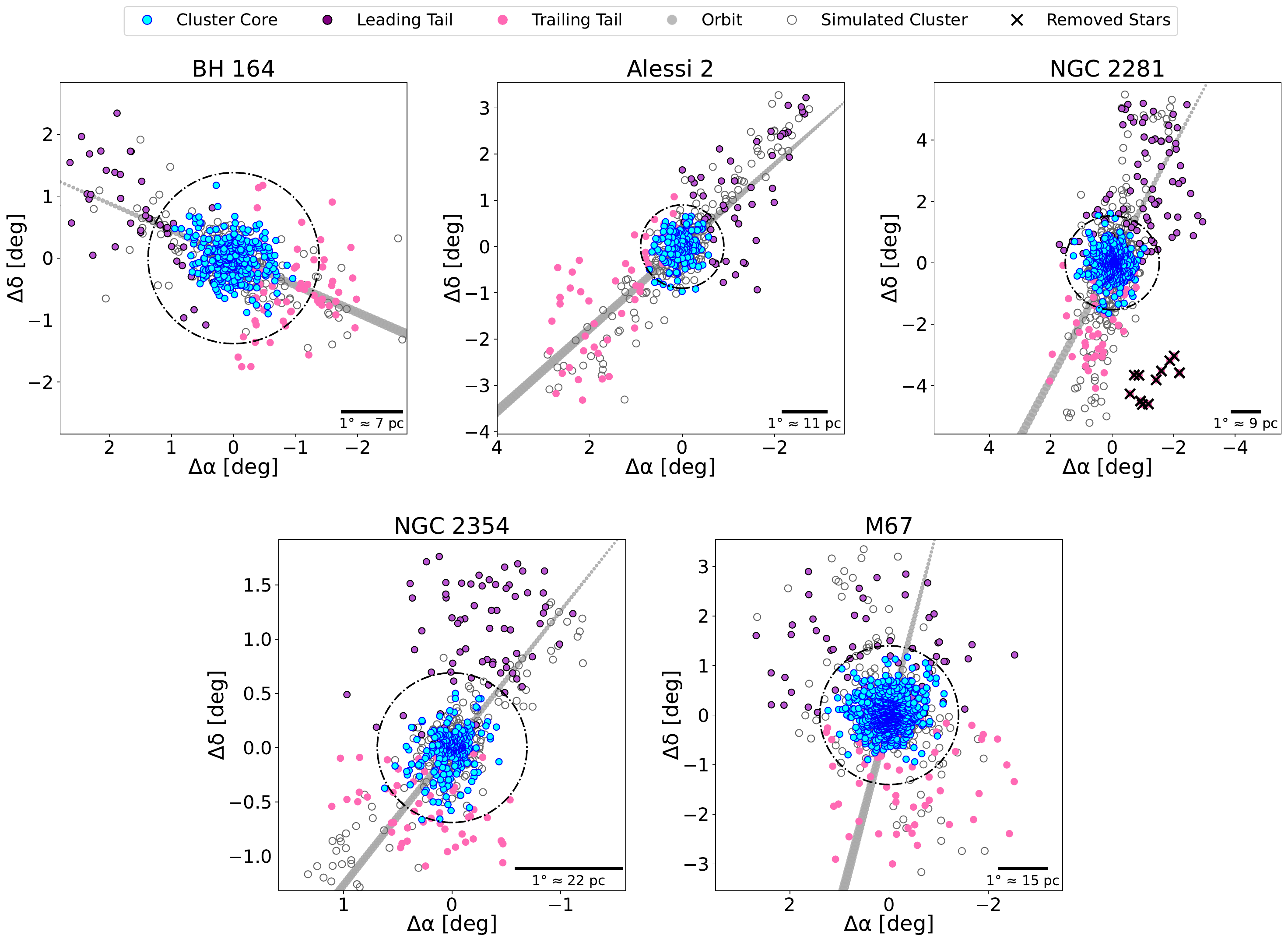} 
\caption{Spatial distribution of core members (blue dots), tidal tails (purple \& pink dots), N-body simulated cluster (empty dots), Jacobi limit (dotted circle), and expected orbit (grey arrow) for the five clusters, with manually removed stars for NGC 2281 (crosses). For details, see \citet{2025arXiv250819457S}.}
\label{fig:2}
\end{figure}

\subsection{DBSCAN Clustering} 
The DBSCAN (Density-Based Spatial Clustering of Applications with Noise) identifies groups of closely packed points while treating outliers as noise. The algorithm was run in the scaled 5D space to identify overdensities corresponding to the core cluster. Parameters ($\epsilon$, min\_neighbors) were tuned for high spatial completeness and low contamination.

\subsection{Tidal Tail Detection} Following the core cluster identification, non-cluster members from the \emph{Filtered Sample} within 500 pc of the mean cluster distance were tested for the tail membership using:
\begin{enumerate}

\item \textbf{Radial Velocity (RV) Filtering}: A $\pm 10$ km/s window around the core cluster's mean radial velocity was applied only on the stars with available RV data as a loose filter to remove clear outliers while preserving plausible candidates. 

\item \textbf{CMD Matching}: KD-Tree structure helped in the efficient nearest-neighbor searches to identify stars with CMD properties similar to core members. A distance threshold of 0.05 in CMD space ensured photometric similarity, with the distance calculated as:
\begin{equation}
d_{\text{CMD}} = \sqrt{(G - G_{\text{core}})^2 + \left[(BP - RP) - (BP_{\text{core}} - RP_{\text{core}})\right]^2}
\end{equation}

Only stars with $d_{\text{CMD}} \leq 0.05$ were retained for further analysis.

\begin{figure}[t]
\centering
\begin{subfigure}[b]{0.49\textwidth}
    \centering
    \includegraphics[width=\textwidth,height=0.26\textheight]{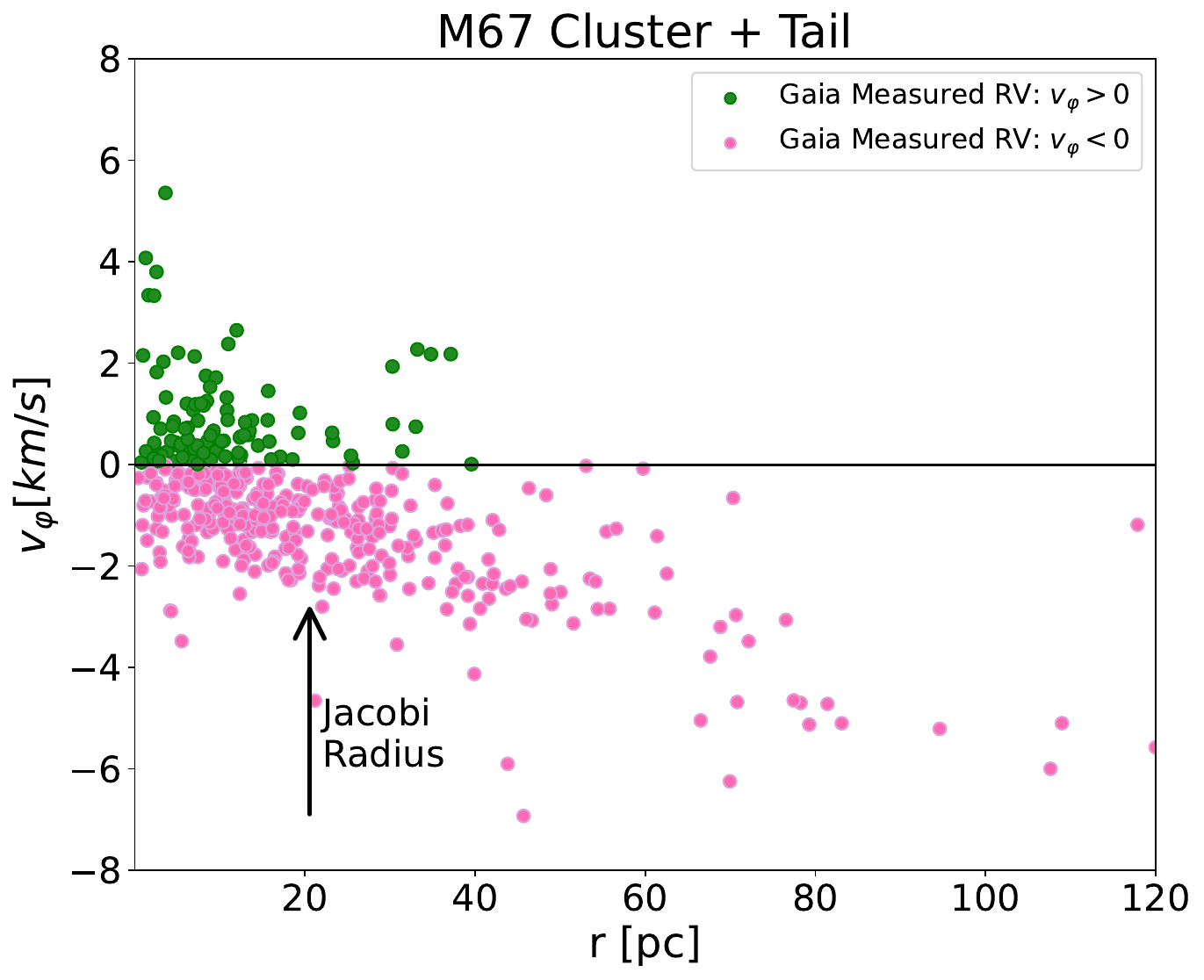}
    \caption{Rotational velocity ($v_\varphi$) vs. distance from M67’s center for cluster + tail stars.}
    \label{fig:2a}
\end{subfigure}
\hfill
\begin{subfigure}[b]{0.49\textwidth}
    \centering
    \includegraphics[width=\textwidth,height=0.26\textheight]{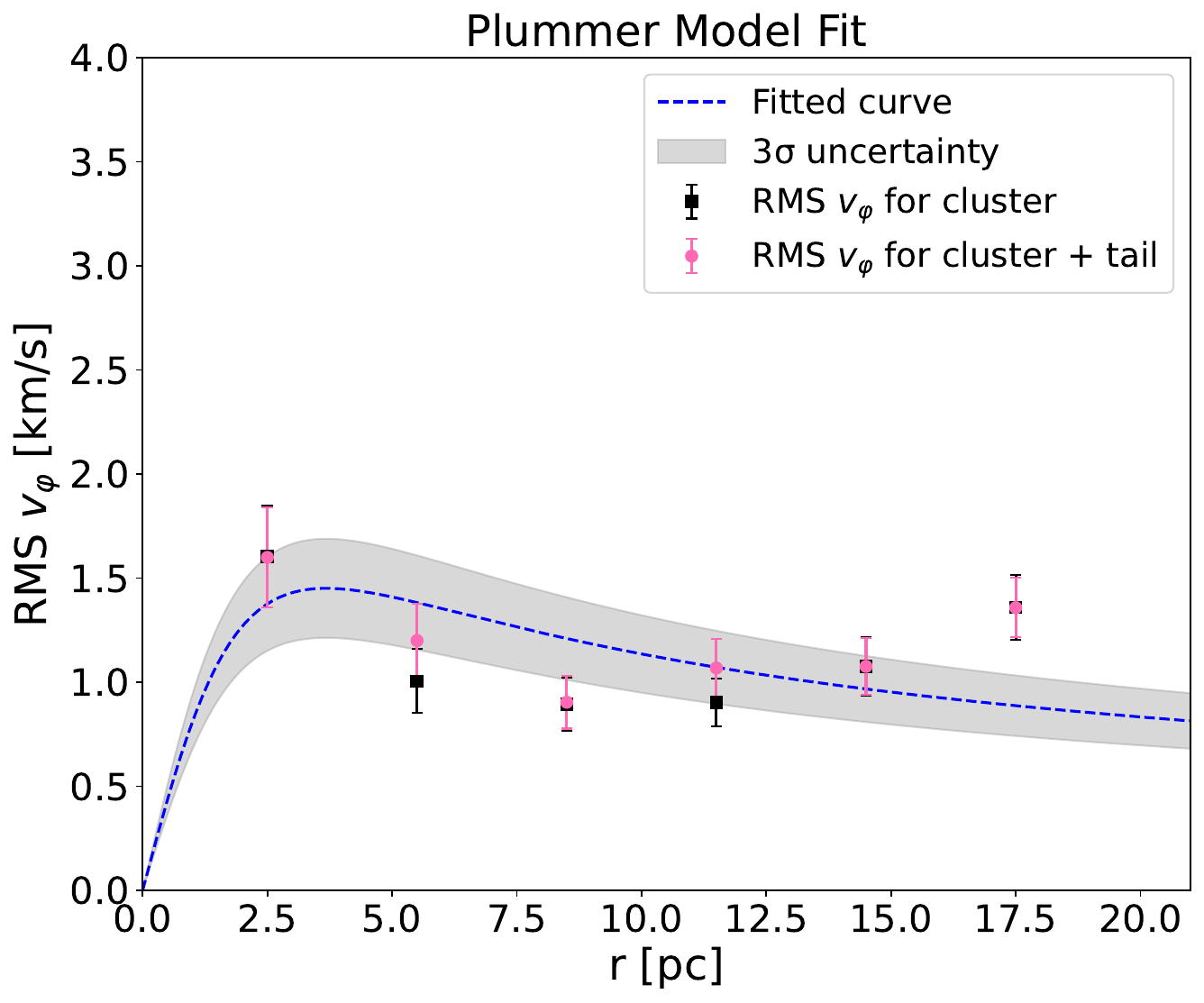}
    \caption{RMS $v_\varphi$ vs. distance from M67’s center with \(3\sigma\) bounds and Plummer fit (core radius 2.6\,pc).}
    \label{fig:2b}
\end{subfigure}
\caption{Rotational analysis of M67 (See Section 3: \autoref{sec:cluster_rotation}).}
\label{fig:rotation_dispersion}
\end{figure}

\item \textbf{Principal Component Analysis (PCA) and Clustering}: PCA was used to determine the cluster's primary elongation axis and align the
coordinate frame accordingly. DBSCAN was then applied to the transformed spatial data to distinguish the tail stars from field noise.

\item \textbf{Tangential Projection of the Celestial Coordinates}: Celestial coordinates were transformed to a 2D tangent plane using the cluster core's mean position as a reference to simplify the visualization and reduce spherical geometry effects.

\item \textbf{Jacobi Radius Criterion}: The detected features were evaluated relative to the best-fit Jacobi radii \citep{2024A&A...686A..42H}, beyond which the stars are prone to tidal stripping.

\end{enumerate}

\section{Tidal Tail Analysis}
We examined the kinematics of the clusters and their tidal tails using proper motion corrections, orbital modelling, luminosity functions, binary statistics, and rotational dynamics.

\begin{enumerate}
    \item \textbf{Projection Effect Corrections}: Following \citet{2009A&A...497..209V}, we fixed projection-induced distortions caused by bulk cluster motion and angular offsets in proper motions and radial velocities.
    
    \item \textbf{Orbit Integration}: Cluster orbits were computed with \texttt{Astropy} and \texttt{Galpy} using the \texttt{MWPotential2014} model, which made it possible to find the leading and trailing tail orientations in the orbital plane.
    
    \item \textbf{Luminosity Functions}: Using 1 mag bins in absolute magnitude ($M_G$), core cluster and tail luminosity functions were constructed. The relation $\log_{10} N \propto a \times M_G$ was used in linear fits, excluding the faintest bins because of observational incompleteness. The distributions of the core and tail stellar populations were compared using the Kolmogorov-Smirnov (KS) test. 
    
    \item \textbf{Binary Fractions}: They were computed by identifying unresolved binaries in Gaia CMDs, using best-fitting PARSEC isochrones \citep{Bressan2012MNRAS.427.127B} based on literature parameters. Main sequence ridge lines were estimated using \textsc{robustgp} \citep{Li2020ApJ...901...49L}, with mass ratios calculated from main sequence shifts. Sources with mass ratios $> 0.5$ were classified as binaries.
    
    \item \textbf{Cluster Rotation}\label{sec:cluster_rotation}: Rotational properties were studied using the residual velocity method \citep{2022ApJ...938..100H, 2024ApJ...963..153H} to identify the rotation axes and compute the azimuthal velocity component $v_{\varphi}$. Analysis was limited to stars with reliable radial velocities ($G < 15$ mag, RV errors $< 5$ km/s), restricting the sample clusters to only M67 and NGC 2281. Radial profiles of $v_{\varphi}$ and RMS $v_{\varphi}$ were examined, with the Plummer model fitted for the mass estimation within the Jacobi radii for both clusters.

\end{enumerate}

\section{Results}
All clusters exhibited tidal tails extending 1–4 times their Jacobi radii (Figure~\ref{fig:2}), with tail lengths of 40–100 pc. The youngest cluster (BH 164, 65 Myr) displayed the shortest tidal tail span, while the oldest (M67, 4 Gyr) exhibited the longest. Luminosity functions revealed a lack of high-mass stars in tails, and binary fractions were higher in tails, potentially due to dynamical interactions ejecting binaries.
Rotation was found in the core and tails of M67 ($v_\varphi = -0.84 \pm 0.07$ km/s), while NGC 2281 showed rotation ($v_\varphi = -0.53 \pm 0.13$ km/s) mostly in its core. Both of them followed the Newtonian dynamics within their Jacobi radii. The Plummer model fits gave mass estimates of \(832\)–\(876\,M_\odot\) for M67 and \(741\)–\(803\,M_\odot\) for NGC 2281.

\begin{table}[h!]
 \centering
 \caption{Core cluster and tidal tail parameters for the five open clusters. Columns 1–3: Name and equatorial coordinates; Columns 4–5: Proper motions; Column 6: Mean RV of core cluster members; Column 7: Number of core members; Column 8: Number of tidal tail stars; Column 9: Best-fitting Jacobi radius from \citet{2024A&A...686A..42H}; Column 10: Total projected tip-to-tip extent of the tidal tails; Column 11-12: Binary fractions in the clusters and tails.}
{\tablefont\begin{tabular}{@{\extracolsep{\fill}}lccccccccccc}
\midrule
Name & RA & DEC & pmRA & pmDEC & RV & $N_{memb}$ & $N_{tail}$ & $r_J$ & $L_{total}$ & $BF_{cluster}$ & $BF_{tail}$ \\
 & (deg) & (deg) & (mas/yr) & (mas/yr) & (km/s) & & & (pc) & (pc) & & \\
\midrule
BH 164 & 222.20 & -66.45 & -7.41 & -10.69 & $-3.4 \pm 3.3$ & 289 & 107 & 9.9 & 40 & $0.26\pm0.04$ & $0.26\pm0.08$ \\
Alessi 2 & 71.61 & 55.16 & -0.92 & -1.08 & $-10.3 \pm 4.2$ & 130 & 100 & 9.6 & 90 & $0.22\pm0.05$ & $0.47\pm0.12$ \\
NGC 2281 & 102.09 & 41.05 & -2.96 & -8.27 & $19.6 \pm 4.6$ & 438 & 136 & 13.6 & 91 & $0.24\pm0.03$ & $0.28\pm0.08$ \\
NGC 2354 & 108.49 & -25.73 & -2.86 & 1.86 & $31.8 \pm 5.5$ & 244 & 142 & 14.9 & 69 & $0.34\pm0.06$ & $0.27\pm0.08$ \\
M67 & 132.85 & 11.82 & -10.97 & -2.92 & $33.9 \pm 3.4$ & 1037 & 163 & 20.6 & 102 & $0.28\pm0.02$ & $0.30\pm0.09$ \\
\midrule
\end{tabular}}
\label{tab:combined_parameters}
\end{table}

\section{Conclusion}
Our results showed that optimized clustering methods can detect extended structures up to a kiloparsec, increasing the range of clusters that can be studied compared to other techniques, such as the convergent point method, which is typically limited to within $\approx$400 pc. Moreover, detecting rotation in outer regions can also be used as a test for validating the presence of tidal tails, as 6D information is poor at larger distances. Extending this approach to more clusters can reveal hidden structures and refine dynamical models of cluster evolution.

\begin{acknowledgements} I.S. acknowledges financial support from the International Astronomical Union through a travel grant to attend IAU Symposium 398 / MODEST-25. 
This work has made use of data from the European Space Agency (ESA) mission Gaia (\href{https://www.cosmos.esa.int/gaia}{https://www.cosmos.esa.int/gaia}), 
processed by the Gaia Data Processing and Analysis Consortium (DPAC, \href{https://www.cosmos.esa.int/web/gaia/dpac/consortium}{https://www.cosmos.esa.int/web/gaia/dpac/consortium}). 
\end{acknowledgements}

\end{document}